\documentclass{aa}
\usepackage{graphicx}
\usepackage{txfonts}
\usepackage{amsmath}
\usepackage{comment}

\newcommand{\RNum}[1]{\uppercase\expandafter{\romannumeral #1\relax}}

\begin{document} 
   \title{The bar resonances and low angular momentum moving groups \\in the Galaxy revealed by stellar ages}
   \titlerunning{Bar resonances and low AM moving groups revealed by age}
\authorrunning{C.F.P. Laporte et al.}
   \author{
    Chervin F. P. Laporte \inst{1,2,3}
    \and
    Benoit Famaey \inst{4}
    \and
    Giacomo Monari \inst{4}
    \and
    Vanessa Hill \inst{2}
    \and
    Christopher Wegg \inst{2}
      \and
    Ortwin Gerhard \inst{5}  
   }
    \institute{Kavli Institute for the Physics and Mathematics of the Universe (WPI), The University of Tokyo Institutes for Advanced Study (UTIAS), The University of Tokyo, Chiba 277-8583, Japan
    \email{chervin.laporte@ipmu.jp}
    \and
    Laboratoire Lagrange, Universit\'e C\^ote d'Azur, Observatoire de la C\^ote d’ Azur, CNRS, Bd de l’Observatoire, 06304 Nice, France
    \and
    Department of Physics \& Astronomy, University of Victoria, 3800 Finnerty Road, Victoria BC, V8P 5C2 Canada 
    \and
    Universit\'e de Strasbourg, CNRS UMR 7550, Observatoire astronomique de Strasbourg, 11 rue de l'Universit\'e, 67000 Strasbourg, France
    \and
    Max-Planck-Institut f\"ur extraterrestrische Physik, Gießenbachstraße 1, 85748 Garching bei M\"unchen, Germany
    }
   \date{Received xxxx; accepted xxxx}
  \abstract
   { We use the second Gaia data release in combination with the catalog of Sanders \& Das (2018) to dissect the Milky Way disc in phase-space and relative ages. We confirm and report the existence of multiple velocity moving groups at low azimuthal velocities and angular momenta, below Arcturus, regularly separated by $\sim~18-20\,\rm{km s^{-1}}$ in azimuthal velocity. Such features were predicted to exist more than ten years ago from the reaction of the Milky Way to a perturbation in the disc undergoing phase-mixing. These structures appear slightly younger than their phase-space surroundings, arguing against an extra-galactic origin. We also identify in relative age many of the classical ridges in the plane of azimuthal velocity vs. Galactocentric radius, traditionally associated with resonance features. These ridges are also younger than their phase-space surroundings in accordance with predictions from recent state-of-the-art cosmological hydrodynamical simulations of Milky Way-like galaxies. We study the response of dynamically young and old stellar disc populations to resonances from an analytic model of a large bar, which, remarkably, qualitatively reproduces the trends seen in the data. Our results re-inforce the idea that the Galactic disc is currently being shaped by both internal and external perturbations, and that, while absolute isochrone ages have to be taken with great care, exploring the dynamical structure of the disc in stellar ages, especially with future asteroseismic data, will provide much stronger constraints than metallicity/abundance trends alone.
   }
   \keywords{Galaxy: kinematics and dynamics -- Galaxy: disc -- Galaxy: solar neighborhood -- Galaxy: structure -- Galaxy: evolution}
   \maketitle
%
\section{Introduction}
The velocity distribution in the solar neighbourhood is far from smooth and has long been noted to be substructured. A number of moving groups have indeed been known for a very long time \citep[e.g.,][]{,eggen1965,dehnen98,famaey05}, with their various origins still debated. Proposed mechanisms range from internal resonances due to the bar or spiral arms \citep[e.g.][]{dehnen99,perez-villegas17,monari19,hunt19, fragkoudi19}, phase-mixing signatures from external perturbations \citep{minchev09, hunt18}, to even extra-galactic accretion origins \citep{navarro04}.

This observational fact recently became even more striking with the Gaia DR2 catalogue \citep{GaiaBrown}, revealing in unprecedented detail prominent arches in local velocity \citep{GaiaKatz} and action space \citep{trick19a}. It is only recently that action-space modeling, despite their strong advocacy, has been used to treat perturbations \citep[e.g.,][]{monari17, binney018, monari19}, which has allowed one to recover many of the known moving groups at azimuthal velocities $v_\phi$ close to the Sun's velocity. However, in part due to debates and uncertainties about the actual pattern speed of the bar \citep{monari17b, perez-villegas17, fragkoudi19, trick19, monari19}, such modeling did not predict or reveal new substructures in the solar neighbourhood. Instead, using unsharp masking on the Gaia DR2 data, \citet[][see their figure 13]{laporte19c} unveiled the existence of low angular momentum ($L_z$) velocity substructures, long ago predicted to exist at azimuthal velocities $v_\phi$ below Arcturus from phase-mixing resulting from an external perturbation \citep{minchev09}. While the regular separation in velocity space in the ridges is a clearly distinct hint of such phase-mixing in the disc, kinematics alone unfortunately does not allow one to posit an in-situ or ex-situ origin, as has been debated for Arcturus. In order to lift any further doubts, these structures need to be studied in a complementary space to that of phase-space. While chemistry, such as metallicity or alpha-abundances, has been used to study the substructured content of the solar vicinity \citep[e.g.][]{khanna19}, the presence of spatial metallicity gradients make interpretative tasks rather difficult. Another issue, especially in the case of Arcturus, is the low number of spectroscopically confirmed candidates with the specific kinematics used in the past, which may give the illusion of a tight chemical sequence suggesting the putative existence of an accreted dwarf galaxy \citep[see e.g.][]{navarro04}. An added-value physical quantity is thus required, which can be provided by differential stellar ages. The recent catalog by \cite{sanders18} provides a compilation of ages derived for about 3 million stars from a combination of legacy spectroscopic surveys with Gaia proper motions, ensuring a full 6-D characterisation. While such isochrone ages have to be treated with great care in absolute terms, they can nevertheless be very useful to identify patterns in term of relative ages. Moreover, differential ages as a function of velocity at a given position cannot be affected by selection effects.

\begin{figure*}
\includegraphics[width=1.0\textwidth,trim=0mm 25mm 0mm 20mm, clip]{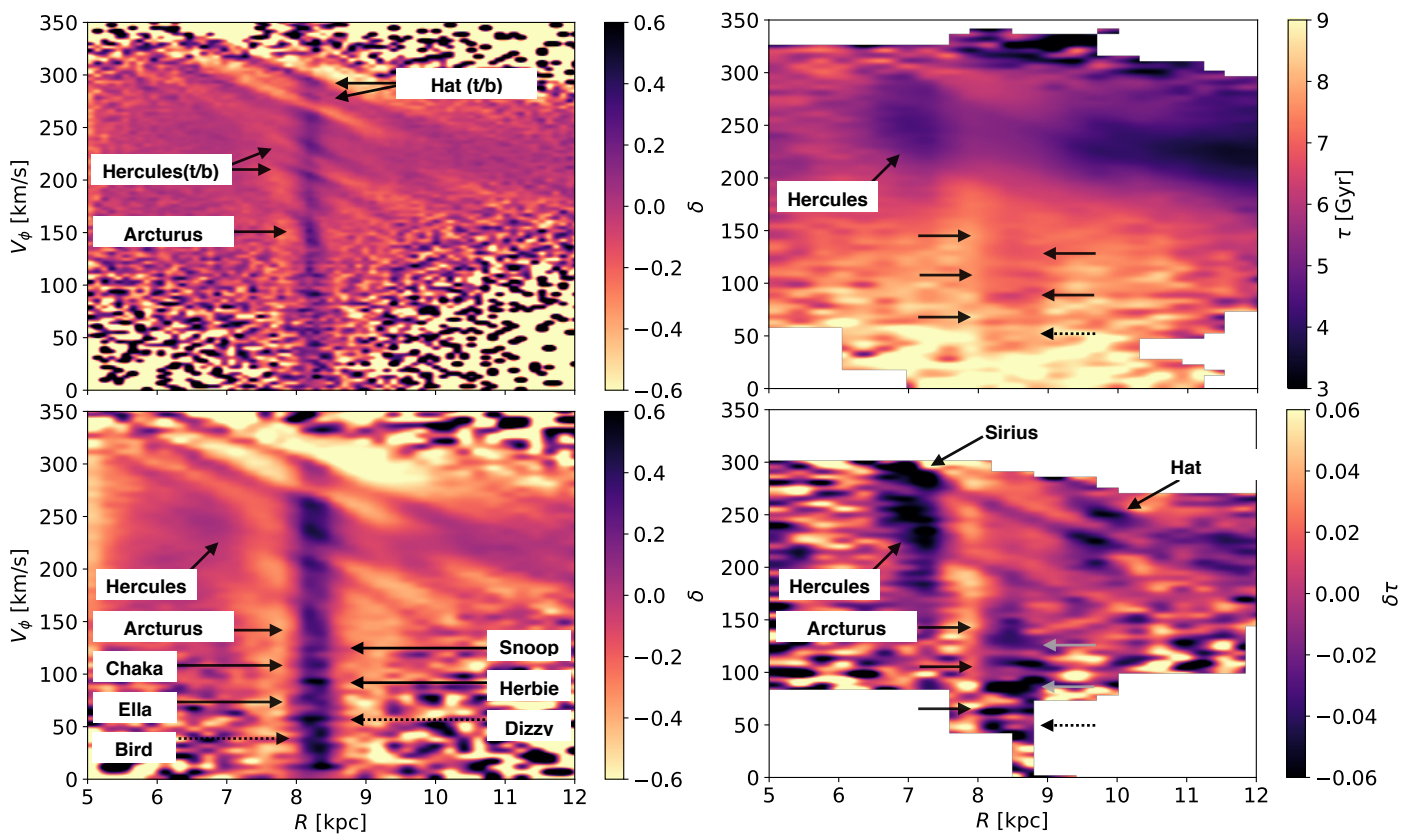}
\caption[]{ {\it Top Left}: Unsharp masked $\delta(v_{\phi}-R)$ plane for the Gaia DR2 and Sanders \& Das (2018) master catalog ($\sim9$ million sources with 6-D phase-space information). The binsize is $(\Delta_R, \Delta v_{\phi})=(0.1\,\rm{kpc}, 3\, \rm{km s^{-1}})$ and helps visualise the finer parts of the classical ridges (Sirius, Hyades, Hercules). We note that the `hat' is split in two (which is expected at the OLR from the orbits trapped at the resonance and the linear deformation of velocity space below the resoance). Arcturus also stands out. {\it Bottom Left:} Similar to the top left figure but with binsize $(\Delta_R, \Delta v_{\phi})=(0.2\,\rm{kpc}, 5\, \rm{km s^{-1}})$ to help low $L_z$ moving groups below Arcturus stand out. These are separated by $\sim18-20\,\rm{km s^{-1}}$ consistent with the Minchev et al. (2009)'s predictions (Snoop, Chaka, Herbie/K08), the series extends further to lower velocities (e.g. `Ella' $v \sim 70 {\rm km s^{-1}}$, `Dizzy' $v\sim50 {\rm km s^{-1}}$, and even perhaps what we denote `Bird' at $\sim 30 {\rm km s^{-1}}$). {\it Top Right:} Age moment $\tau$ of the $v_{\phi}-R$ plane for the Sanders \& Das (2018) catalog. Ridges from Hercules, Sirius and the `hat' (OLR) appear strongly. low $L_z$ moving groups (Arcturus, Snoop, Chaka, Herbie, Ella) also show up as slightly younger patches. We caution against interpreting those absolute ages as a function of radius, as selection effects might be important. What is more useful is to compare ages as a function of velocities at a fixed radius. {\it Bottom Right:} Differential age ($\delta\tau=\tau/\bar{\tau}-1$) distribution in the $v_{\phi}-R$ plane. Low $L_z$ ridges show up as slightly but prominently younger than their surrounding. The `hat' (OLR) shows a prominent signal at radii beyond the Sun ($9<R/\rm{kpc}<11$). Sirius is present across the whole range ($7<R/\rm{kpc}<11$), and Hercules is particularly strong at radii smaller than $R_0$ with an extension out to $9.5\rm{kpc}$. }
\end{figure*}

Simulations of disc galaxies stratified as a superposition of different stellar age populations, and as a result with different kinematics through the age-velocity dispersion relation \citep[AVR, see, e.g., figure 9 of][for observations]{miglio20}, show that younger, colder disc populations react more strongly than older hotter components, whether subjected to internal perturbations \citep[e.g.,][]{fragkoudi19} or satellite impacts \citep[e.g.,][]{chequers18}. Thus characterising the $v_{\phi}-R$ plane in stellar ages would prove particularly useful to study the possible disc or accreted origin of low $L_z$ moving groups, from Arcturus to lower azimuthal velocities. Moreover, it would also help isolate the signatures of bar resonances in the disc which would manifest themselves at distinct radii depending on the pattern speed and intrinsic structure of the Galactic bar. Typical modeling efforts to study bar resonances in the solar neighbourhood, starting with \citet{dehnen99}, have represented the bar as a simple m=2 quadrupole with varying fixed pattern speeds. However, taking into account higher-order modes in the Fourier decomposition of the bar, \cite{monari19} have shown that the situation is more complex.

In this letter, we study the  $v_{\phi}-R$ plane in the space of stellar ages to decipher the origin of low $L_z$ moving groups as well as the resonances of the bar in the Galaxy as predicted by an analytical action-angle perturbative model based on the large bar of \citet{portail17}, already used in \citet{monari19}. In section 2 we present the data used in this study, and the actual $v_\phi-R$ plane as revealed through stellar ages. In section 3 we compare these results to an analytical bar model to interpret some age structure seen in the data. In section 4, we (re-)analyze the chemical structure of low $L_z$ moving groups, before concluding in section 5.

\section{Data and age dissection of phase-space}

The data we use for our analysis comes from the publicly available catalog of \citet{sanders18} which combines Gaia DR2 data with various current legacy surveys in order to derive isochrone ages for $\sim3$ million stars. We refer the reader to the original paper for more details. In Fig.~1, we present a series of $v_{\phi}-R$ planes in overdensity (top/bottom left panels), age and differential ages (top and bottom right panels respectively). For the density plots, we complement the \citet{sanders18} catalog with all the Gaia DR2 stars with measured line-of-sight velocities and $\varpi/\sigma\varpi>5$ (RVS). For the RVS stars, we use the distances derived from \cite{bailerjones18}. Given our focus on structures relatively close to the solar neighbourhood, changing to other distance estimates did not change any of our conclusions. For consistency with \citet{sanders18}, we transformed from ICRS to Galactocentric coordinates by adopting the position of the Sun at $R_0 = 8.2\,\rm{kpc}$,  a local circular velocity of $v_c(R_0) = 240\,\rm{ km s^{-1}}$ and peculiar Solar motion with respect to the Local Standard of Rest of $(U_{\odot} , V_{\odot} , W_{\odot}) = (11.10, 12.24, 7.25) \,\rm{km s^{-1}}$ \citep{schoenrich10} and a height of the Sun about the midplane of $z_{\odot}=15\,\rm{pc}$. Together, this gives us a master catalog of $\sim 9$ million stars with full 6D phase-space to probe the $v_{\phi}-R$ plane \footnote{We checked that both catalogs separately give the same features in the extent/position of the classical ridges in terms of density and velocity moments as explored in \citet[][see their Fig. 13]{laporte19c}, but the combined catalog increases the sampling, particularly in the solar neighbourhood.}. 

The presence of the main dominant classical ridges, revealed in number counts by Gaia DR2 \citep{GaiaKatz, Kawata18, antoja18, ramos18}, including Hercules and the `hat', is evident in Fig.~1, together with the additional lower velocity ridges below Arcturus, all separated by $\sim18-20 km s^{-1}$ in density ($\delta=\rho/\bar{\rho}-1$) but also in the age moment map of the $v_{\phi}-R$ plane (denoted by the black arrows in Fig.~1). 
Two of them (denoted `Snoop' and `Chaka') were predicted by \citet{minchev09} to lie between Arcturus and `Herbie'/K08\footnote{Herbie was in fact uncovered using phase-space information alone using the first RAVE catalog by \cite{klement08}.}. We also note the presence of a few low $L_z$ moving groups further below Herbie/K08 (Fig.1 bottow left) such as the ridges at $v_{\phi}\sim70, 50, 30\, \rm{km s^{-1}}$ (which we label "Ella", "Dizzy", "Bird"), digging deeper in the halo-dominated region of velocity-space. Some of these ridges were tentatively seen in \citet{klement08} however the authors did not further comment on their result due to the paucity of tracers used at the time. While the Gaia DR2 sample increased this, given the little complementary data available (age/metallicity/abundance) we leave them as curiosities (hence the dashed arrows in Fig. 1) to be explored in detail with future Data Releases. 

Although low $L_z$ moving groups such as Arcturus have been proposed to originate from an extra-galactic origin, we see that the ridges from Arcuturus down to Ella all have stellar ages that are systematically younger than the surrounding population. This is clear evidence against an extra-galactic origin for such structures, which otherwise should have been, if anything, older than the surroundings. Combined with the regularity in intervals between ridges in velocity space, this clearly argues for a relic sign of a perturbation to the Milky Way disc which is currently undergoing phase-mixing \citep{minchev09}. Proposed scenarios include internal perturbations from transient spiral arms \citep{hunt19}, a slowing bar \citep{chiba}, or external perturbations by a satellite which naturally seed spirals and give rise to ridges \citep{purcell11,gomez12, gomez13,laporte18b, laporte19c}. Disentangling between these processes acting on Galactic discs which may act simultaneously likely requires development of tools beyond those traditionally used (test-particles/N-body simulations) and are beyond the scope of this letter.  
. 

Amongst the classical ridges, we notice the `hat' \citep[the L2 ridge from][]{ramos18}  - identified as the outer Lindblad resonance (OLR) in our bar model in the next section - which is split in density as already noted in \citet{laporte19c} (see their Figure 13 top panel) with an outward/inward radial motion split \citep{laporte19c, fragkoudi19}. Then follow, from large to lower $L_z$, Sirius, Hyades, and Hercules which we further analyse. In the space of relative ages, Sirius is very prominent through a wide range of radii ($6<R/\rm{kpc}<12$). In contrast, the `hat' stands out most strongly at radii beyond $R\sim9\,\rm{kpc}$ and Hercules appears (albeit more weakly than Sirius) in the solar neighbourhood and picks up more strongly towards $R\sim7\,\rm{kpc}$, both in terms of differential ages and in the global density in the $v_\phi-R$ plane. These interesting trends are best appreciated in the differential age map of the $v_{\phi}-R$ plane.

\begin{figure}
\includegraphics[width=\columnwidth]{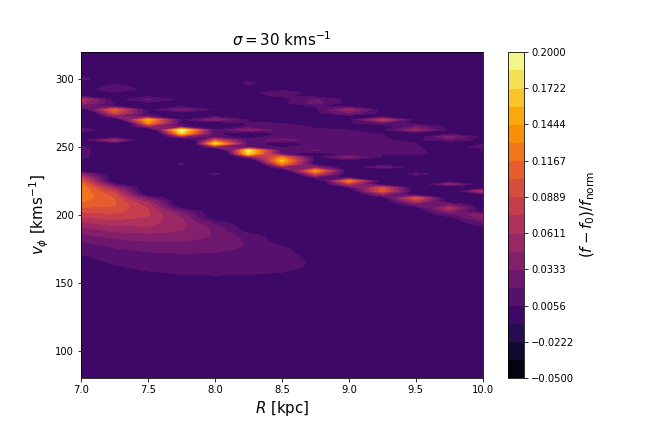}
\includegraphics[width=\columnwidth]{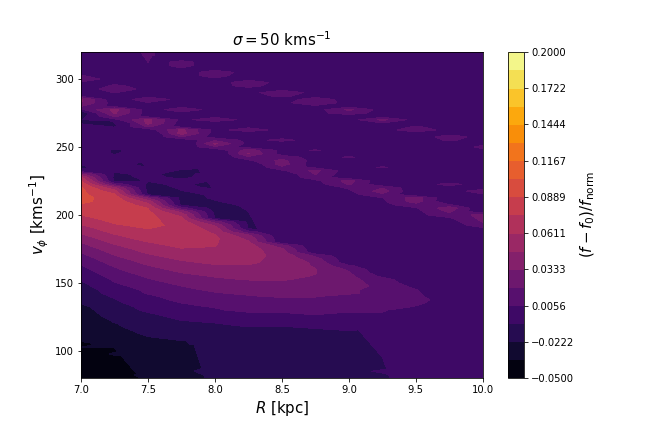}
\caption[]{The normalized perturbed distribution function in the presence of the large bar perturbation of \citet{portail17}. {\it Top panel}: for an unperturbed distribution function representing a young and cold disc stellar population with $\sigma_R=30 \, {\rm km} \, {\rm s}^{-1}$ at the Galactocentric solar radius $R_0$. {\it Bottom panel}: for an older and hotter disc population with $\sigma_R=50 \, {\rm km} \, {\rm s}^{-1}$ at the Galactocentric solar radius $R_0$.}
\end{figure}

\section{Analytic bar model}

In this section, we consider an analytic model as developed by \cite{monari17} aimed at studying the response of a disc composed of a dynamically young (low velocity dispersion) and old (high velocity dispersion) stellar populations to the large bar model of \citet{portail17} with pattern speed $\Omega_{b}=39\,\rm{km s^{-1}/kpc}$. This dynamical model is based on the number density of red clump stars in the bulge and long bar \citep[c.f.][]{wegg15} and the line-of-sight velocities of large samples of bulge
giants. Already in \citet{monari19}, this model was shown to reproduce a number of ridges in the Gaia DR2 data which they associated with the OLR, corotation (CR), as well as the 4:1 and 6:1 resonances from higher order Fourier modes of the bar potential. 

We define a quasi-isothermal \citep{binney10} unperturbed distribution function in action space to represent a young and cold  ($\sigma_R=30 \, {\rm km} \, {\rm s}^{-1}$), as well as a hotter and older ($\sigma_R=50 \, {\rm km} \, {\rm s}^{-1}$) disc stellar population. Then we use a first canonical transformation to move to the slow and fast action-angle variables, which allows to determine the regions of phase-space where orbits are trapped to resonances. In the vicinity of these resonant regions, and after averaging over the fast angles, the dynamics of the slow variables is that of a pendulum, for which we make a new canonical transformation to pendulum action-angle variables. We then phase-mix the original unperturbed distribution function over the pendulum angles to obtain the perturbed distribution function in the different resonant zones. We can then compute the perturbed distribution function at different positions and velocities, making use of the epicyclic approximation. We refer the reader to \cite{monari17,monari19} for more details of the method. Note that it gives a very good representation of orbits trapped at resonances, but is not the best possible description of the linear deformation zones. In Fig.~2, we present the variations in density of the $v_{\phi}-R$ plane resulting from the response of both the cold (young) and hot (old) stellar populations to the bar potential of \citet{portail17}. As expected, the features are marked much more clearly for the dynamically young population. Interestingly, this is particularly true for the 4:1 resonance, which corresponds to the Sirius ridge, which is also seen very prominently in the data as a young feature in a large range of radii. Moreover, the density contrast of the ridge associated with corotation is most prominent for young stars around $R\sim7$~kpc, which is also where the Hercules ridge appears most clearly both in global density and as a young feature in the data. The `hat' feature, which corresponds to the bar's OLR in this model appears prominently for the young and cold population at $R\sim9$~kpc, where it also peaks in differential age in the data. This analytical model thus provides a remarkable qualitative match to the trend uncovered in the data.

\section{Arcturus and moving groups of the thick disc}

\begin{figure}
\includegraphics[width=\columnwidth]{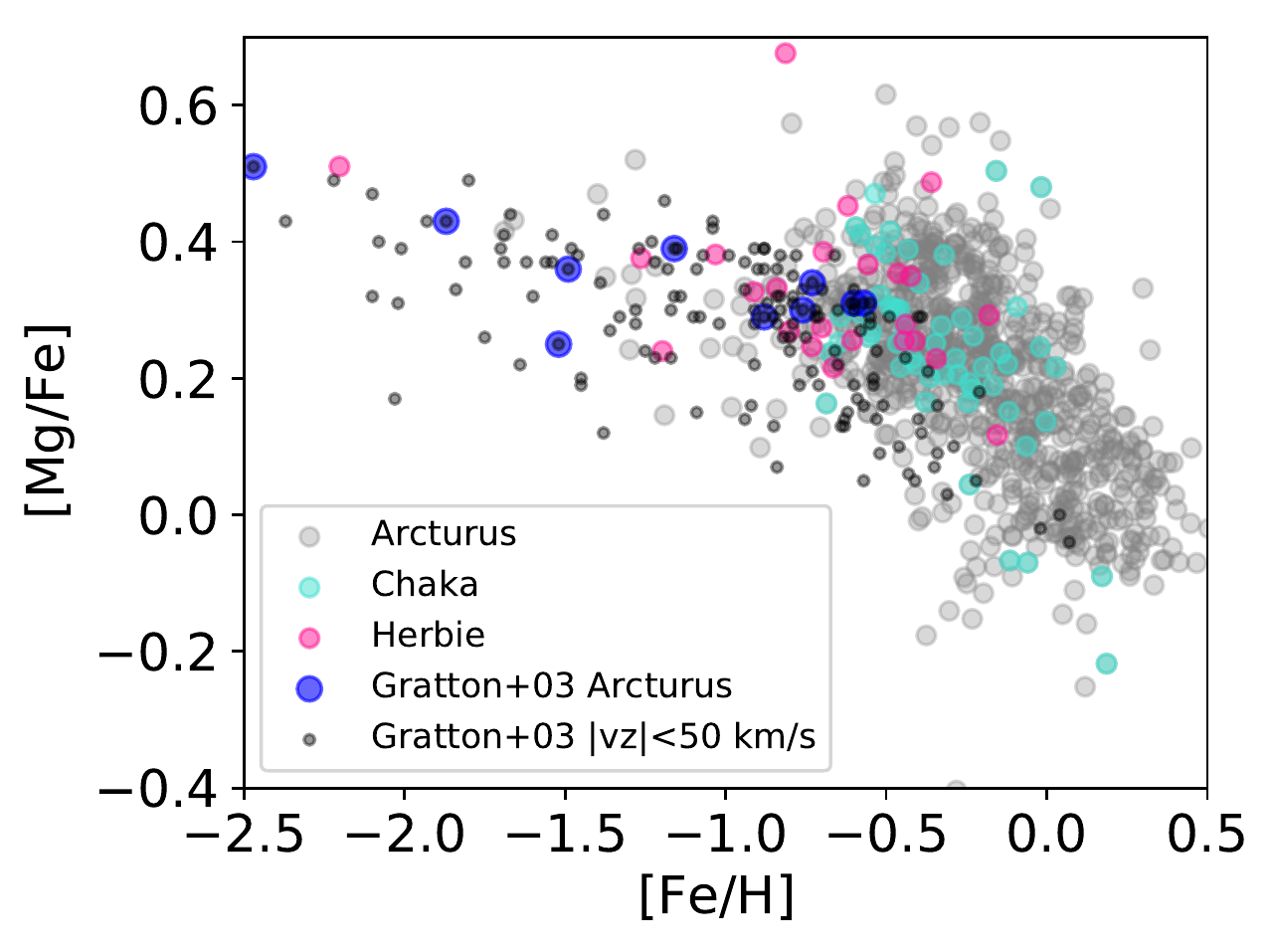}
\caption[]{[Mg/Fe] vs [Fe/H] plot for stars in Arcturus (as selected by N04 - which intrisincally includes Snoop), Chaka and Herbie from APOGEE in thick grey, turquoise and magenta dots. We also overlay the original stars used in N04 from the GCCLB dataset. We note that this dataset of metal poor stars was biased and falls at the boundary with the disc as well as the continuation of the chemically thick disc. The larger APOGEE DR14 sample extends just as much into the metal poor end probed by GCCLB and together with the N04 Arcturus sample gives a more complete census on the origin of Arcturus from the chemical side alone.}
\end{figure}

We had already noted in section 2 that the ridges from Arcuturus down to Ella all have stellar ages that appear slightly but systematically younger than the surrounding population, which, combined with the regularity in intervals between ridges in velocity space, argues for a relic sign of a perturbation to the Milky Way disc. In this section, we now use chemistry to confirm this picture in the case of Arcturus and low $L_z$ moving groups. 

Using the catalog of metal-poor stars from \citet[hereafter GCCLB][]{Gratton03},  \citet[][hereafter N04]{navarro04} selected stars in the solar neighbourhood from the Arcturus group following two criteria:  $1100<L_{z}/ \rm{km s^{-1}\,kpc} <1300$ \& $|v_{z}|<50 \,\rm{km s^{-1}}$. They found a tight sequence in the ([$\alpha/$Fe], [Fe/H]) plane w.r.t. the full GCCLB sample which they used as an argument for a possible extra-galactic origin. However, it must be noted that the sample used was modest in size and biased towards low metallicities, which stop around [Fe/H]$\sim-0.6$. As we shall see, such biases easily mimick a tight sequence. Using the APOGEE DR14 catalog \citep{abolfathi18}, we carry out the same analysis as in N04. In Figure 3, we present the ([Mg/Fe], [Fe/H]) plane for Arcturus and the low $L_z$ moving groups These were selected as follows in the APOGEE catalog:

\begin{itemize}
    \item $d<0.4\,\rm{kpc}$ 
    \item $|z|<1\,\rm{kpc}$ 
    \item $1100<L_{z}/ \rm{km s^{-1}\,kpc} <1300$ \& $|v_{z}|<50 \,\rm{km s^{-1}}$, for Arcturus
    \item $952 < L_{z}/ \rm{km s^{-1}\,kpc} < 929$, for Chaka
    \item $624 < L_{z}/ \rm{km s^{-1}\,kpc} < 860$, for Herbie
\end{itemize}

We note that APOGEE also finds stars in the same region probed by N04, however the Arcturus magnesium sequence is no longer tight but rather (1) broad, (2) with no signs of a knee, and (3) extends well into the thick disc with a contribution from the chemical thin disc. The other low $L_z$ moving groups\footnote{The broad angular momentum selection in N04 for Arcturus also comprises Snoop which we do not show for clarity in Fig. 3.} (e.g. Chaka, Herbie) on the other hand are predominantly composed of thick disc stars and occupy the same space as Arcturus. 

While chemistry has also been used as an argument in favour for a internal origin of the structure, these came from separate analyses \citep[e.g][]{kushniruk19} which did not compare to the original N04 results to understand the differing conclusions. While chemistry is a useful quantity to study, it should still be supplemented by stellar ages as a third complementary and final check. Here, we have shown that the ages of those low $L_z$ moving groups tend to be slightly lower relative to their surroundings. Indeed, even the thick disk displays an AVR \citep[][]{miglio20}. While the association between phase-space structures and younger structures in terms of relative age is a striking result, we caution against the use of isochrone ages to evaluate the {\it absolute} age of stars in these structures. The latter will need future asteroseismic data for large samples in order to quantify the exact age of these structures \citep[][]{miglio20}.




\section{Discussion \& Conclusions}

Identifying signatures associated with the bar in the solar neighbourhood in phase-space alone is notoriously difficult \citep[see][]{trick19}. Other diagnostics such as metallicity, alpha-abundances and ages have been proposed as a way of identifying prominent ridges associated with the OLR for example \citep[e.g.][]{fragkoudi20}. These have been investigated in GALAH for metallicity and abundances \citep{khanna19}. However, stellar ages remain the most physically interesting quantity (albeit less precisely measured) both in terms of probing the growth history of the disc and its correlation with the dynamics through the age-velocity dispersion relation (AVR) \citep[e.g.][]{sanders18, miglio20}. 
 
Here we explore, for the first time, the $v_{\phi}-R$ plane in the space of ages. Interestingly, our findings highlight and confirm predictions from state-of-the-art simulations of Milky Way-like systems that regions associated with resonances of the bar should on average be younger than the surrounding phase-space \citep{fragkoudi20}. We compare the data to predictions from an analytic model based on the large bar of \citet{portail17}, as in \citet{monari19}. The model is able to qualitatively predict the spatial variations and locations of differential age peaks of many known ridges associated with bar resonances such as the `hat' (OLR), Sirius (4:1 resonance) and Hercules (CR). 

Although this letter is a first exploration of the signature of the bar in different stellar age populations, we note a few caveats. We have assumed that the bar has a constant pattern speed with the parameters derived by \citet{portail17}. We have left out the contribution of spiral arms and/or the possibility that the bar is currently reconnecting with a spiral arm and displays fast pattern speed variations \citep{petersen19, hilmi20}. It will be interesting to investigate how such models with varying pattern speed compare to the data presented here.

Another novel result from analysing the $v_{\phi}-R$ plane in differential ages is the complementary confirmation of low $L_z$ (i.e., high-velocity w.r.t. the Sun) moving groups predicted by a model of \citet{minchev09} below Arcturus: Snoop, Chaka, Herbie/K08 and Ella -- the latter two having been previously noted in \citet{klement08}, but at the time not as a significant detection in the case of Ella (only two stars). The association we found between phase-space structures and slightly younger structures in terms of relative age is striking, and argues for a relic sign of a perturbation to the Milky Way disc, especially when combined with the regularity in intervals between ridges in velocity space.

Although Arcturus has been argued in the past to be a substructure associated with an extra-galactic orign (N04), this interpretation was challenged in \citet{minchev09} and lended further credence with Gaia DR2 in \citet{laporte19c}, and also corroborated by \citet{kushniruk19}. Here we provide the final argument in favour of an internal origin based on a re-analysis of the samples used in N04 complemented with APOGEE data \citep{abolfathi18} to understand those starkly different (yet sound) conclusions at earlier times. We show in particular that the sequence argued in N04 does not show a sharp knee, but is broader than that highlighted using only GCCLB, and continues well into the thick disc regimes at [Fe/H]$\sim -0.6$ for similarly kinematically selected stars as in N04 (complementing the GCCLB sample). This conclusion holds as well for the other lower $L_z$ moving groups. 

As an outlook into the future, Gaia DR3 will provide radial velocities and stellar parameters for about 40 million stars and this number is expected to increase with surveys such as WEAVE, SDSS-V and 4MOST, allowing for better decomposition of the phase-space structure of the disc (as well as in metallicity and $\alpha$-elemental abundances). Together with better stellar ages, this should hopefully enable us to better understand the interplay between external and internal perturbations in shaping the disc of the Galaxy and constrain the structure of the central region of the Milky Way. It is already clear that asteroseismology will play a major role in this endeavour \citep{miglio17, huber19, miglio20}.

\begin{acknowledgements}
CL thanks the Observatoire de la C\^ote d'Azur for hosting him and J.F. Navarro for discussions. This work was supported in part by World Premier International Research Center Initia- tive (WPI Initiative), MEXT, Japan. BF and GM acknowledge funding from the Agence Nationale de la Recherche (ANR project ANR-18-CE31-0006 and ANR-19-CE31-0017) and from the European Research Council (ERC) under the European Union's Horizon 2020 research and innovation programme (grant agreement No. 834148). This work has made use of data from the European Space Agency (ESA) mission {\it Gaia} (\url{https://www.cosmos.esa.int/gaia}), processed by the {\it Gaia} Data Processing and Analysis Consortium (DPAC,
\url{https://www.cosmos.esa.int/web/gaia/dpac/consortium}). 

\end{acknowledgements}


\bibliographystyle{aa}
\bibliography{master2.bib}

\end{document}